\documentclass[11pt,a4paper]{article}
\usepackage[utf8]{inputenc}
\usepackage{amsmath}
\usepackage{amsfonts}
\usepackage{amssymb}
\usepackage[dvipsnames]{xcolor}
\usepackage{hep}
\usepackage{color}

\usepackage{mathtools}
\usepackage{jheppub}

\usepackage{ulem}

\providecommand{\f}[2]{\frac{{#1}}{{#2}}}
\definecolor{KBFIred}{RGB}{163,35,47}

\newcommand{\KBFI}{Laboratory of High Energy and Computational Physics, National Institute of Chemical Physics and Biophysics, R\"avala pst. 10, 10143 Tallinn, Estonia.}

\newcommand{\Hend}{H_{\rm end}}

\newcommand{\Hosc}{H_{\rm osc}}

\newcommand{\hkin}{h_{\rm kin}}

\providecommand{\f}[2]{\frac{{#1}}{{#2}}}
\newcommand{\mpl}{M_{\rm Pl}}

\newcommand{\LG}{\mathcal{L}}

\newcommand{\pd}{\partial} 
\newcommand{\td}{\mathrm{d}}
\newcommand{\modu}[1]{\left\vert{#1}\right\vert}
\newcommand{\hc}{\text{ H.c. }}

\newcommand{\rnc}{N_{\rm R}^{\rm c}}
\newcommand{\brnc}{\overline{N^{\rm c}_{\rm R}}}
\newcommand{\rn}{N_{\rm R}}
\newcommand{\brn}{\overline{N}_{\rm R}}
\newcommand{\be}{\begin{equation}} 
\newcommand{\ee}{\end{equation}}

\title{\bf \textcolor{KBFIred}
	{Novel mechanism for CMB modulation in the Standard Model }}

\author[a]{Alexandros Karam}

\author[a,b]{\!\!, Tommi Markkanen}

\author[a]{\!\!, Luca Marzola}

\author[c,b]{\!\!, Sami Nurmi}

\author[a]{\!\!, Martti Raidal}

\author[d]{and Arttu Rajantie}
\affiliation[a]{\KBFI}
\affiliation[b]{Helsinki Institute of Physics, P.O. Box 64, FIN-00014 University of Helsinki, Finland}
\affiliation[c]{Department of Physics, University of Jyv\"askyl\"a, P.O. Box 35, FI-40014 University of Jyv\"askyl\"a, Finland}
\affiliation[d]{Department of Physics, Imperial College London, London, SW7 2AZ, United Kingdom}

\emailAdd{alexandros.karam@kbfi.ee}
\emailAdd{tommi.markkanen@kbfi.ee}
\emailAdd{luca.marzola@cern.ch}
\emailAdd{sami.t.nurmi@jyu.fi}
\emailAdd{martti.raidal@cern.ch}
\emailAdd{a.rajantie@imperial.ac.uk}

\abstract{We demonstrate that light spectator fields in their equilibrium can source sizeable CMB anisotropies through modulated reheating even in the absence of direct couplings to the inflaton. The effect arises when the phase space of the inflaton decay is modulated by the spectator which generates masses for the decay products. We call the mechanism \textit{indirect modulation} and using the stochastic eigenvalue expansion show that it can source perturbations even four orders of magnitude larger than the observed. Importantly, the indirect mechanism is present in the Standard Model extended with right-handed neutrinos. For a minimally coupled Higgs boson this leads to a novel lower bound on the quartic coupling and constrains the neutrino Yukawas below unity.}

\begin{document}

\maketitle

\section{Introduction} 
\label{sec:Introduction}

During the inflationary epoch, scalar fields characterized by mass scales well below the Hubble rate exhibit large-scale fluctuations. When these fields do not take part in driving the inflationary expansion, they are commonly referred to as \textit{spectator fields}. 

The behaviour of spectator fields in de Sitter space has been analysed in Refs.~\cite{Chernikov:1968zm,Dowker:1975tf,Bunch:1978yq,Birrell:1982ix,Linde:1982uu,Allen:1985ux,Allen:1987tz}. 
Describing the behaviour of a light and interacting field in de Sitter space often requires field theory methods beyond the usual perturbative approach~\cite{Hu:1986cv,Boyanovsky:2005sh,Serreau:2011fu,Herranen:2013raa,Gautier:2013aoa,Gautier:2015pca,Tokuda:2017fdh,Arai:2011dd,Guilleux:2015pma,Prokopec:2017vxx,Moreau:2018ena,Moreau:2018lmz,LopezNacir:2019ord}. Alternatively, it is possible to employ the classical stochastic methods laid out in Refs.~\cite{Starobinsky:1986fx,Starobinsky:1994bd} to derive a one-point equilibrium probability distribution of the spectator field values. Arbitrary two-point correlation functions, as well as power spectra, are obtained via a spectral expansion~\cite{Markkanen:2019kpv,Markkanen:2020bfc}. The former define the size of the domains into which a spectator field fragments, with each domain characterized by a coherent field value drawn from the equilibrium distribution. Recent works employing the stochastic formalism include Refs.~\cite{Rigopoulos:2016oko,Tokuda:2018eqs,Cruces:2018cvq,Glavan:2017jye,Hardwick:2017fjo,Vennin:2015hra,Moss:2016uix,Grain:2017dqa,Firouzjahi:2018vet,Pinol:2018euk,Hardwick:2019uex,Fumagalli:2019ohr,Jain:2019wxo,Moreau:2019jpn,Pattison:2019hef,Prokopec:2019srf,Franciolini:2018ebs}.

In modulated reheating scenarios spatial modulations of the inflaton decay width affect the local duration of the reheating process, 
which sources curvature perturbations~\cite{Kofman:2003nx,Dvali:2003em}. 
The modulated reheating scenario has been widely explored in different set-ups including both direct and indirect couplings between the inflaton and the spectator. See e.g.~\cite{Ichikawa:2008ne} for the general formalism and~\cite{Kobayashi:2011hp,Fujita:2016vfj,Lu:2019tjj,Choi:2012cp, DeSimone:2012gq,Cai:2013caa} for models connected to the Higgs field. See also~\cite{Chambers:2007se,Chambers:2008gu,Chambers:2009ki,Fujita:2013bka,Fujita:2014hha} for related scenarios.
Most works employ the mean field approach which works well when the spectator is displaced far from the equilibrium during inflation. 


As shown in Ref.~\cite{Markkanen:2019kpv}, when the spectator is in its equilibrium a full stochastic approach is required for reliable analysis and the outcome may substantially differ from the mean field results. In this work we demonstrate that light spectator fields in their equilibrium generally induce significant modulation of the Cosmic Microwave Background (CMB) even in absence of any direct coupling to the inflaton field. For definiteness, we call this mechanism \textit{indirect modulation}. As the spectator field acquires fluctuations comparable to or exceeding the inflaton mass scale, these interactions induce field dependent effective masses that can kinematically block the inflaton decay channels \cite{Fujita:2016vfj,Lu:2019tjj}. Because of spectator fluctuations, the kinematic blocking is released at different times at different locations, resulting in a spatial modulation of the reheating temperature. 

To introduce the stochastic approach in the indirect modulation mechanism, we first consider a simple model consisting of an inflaton field, a spectator field and a fermion, analyzing the consequences of the indirect modulation mechanism. 
We then analyze the case of the standard model (SM) of particle physics extended with right-handed neutrinos, which provide a suitable decay channel for the inflaton and where the Higgs boson plays the role of light spectator field. As a result, the indirect modulation mechanism constitutes a novel way in which CMB observations can constrain the SM physics through the reheating dynamics. 

The paper is organized as follows: in Sec.~\ref{sec:spec101}, we sketch the stochastic treatment of spectator fields.
In Sec.~\ref{sec:dn} we make use of the $\delta N$ formalism to compute the power spectrum of curvature perturbations. The implementation of the stochastic approach to the indirect modulation mechanism is addressed in Sec.~\ref{sec:toy2}, where we apply it to a phenomenological Yukawa model. Section~\ref{sec:higgs} analyzes the case of the SM, demonstrating that all ingredients required for indirect modulation
are present once neutrino phenomenology is addressed. Finally, we present our results in Sec.~\ref{sec:Results} and conclude by summarizing our work in Sec.~\ref{sec:summary}.
\section{Spectator Fields}
\label{sec:spec101}
We start by providing an introduction to the physics of spectator fields in the stochastic approach. For more details we refer the reader to the Refs.~\cite{Markkanen:2019kpv, Markkanen:2020bfc}.

The fluctuations of a light scalar field $h$ in de Sitter space can be shown to obey a Fokker-Planck equation, equivalent to the following eigenvalue problem:
\begin{align}
	\label{eq:EIGS}
	\left[\frac 12 \left(\frac{\pd^2}{\pd h^2}-v'( h)^2 + v''( h)\right)
	+
	\frac{4\pi^2\Lambda_n}{H^3}\right]
	\psi_n( h)
	=0\,.
\end{align}
Here $H$ is the Hubble rate, $v:=\frac{4\pi^2}{3H^4}V( h)$, $V(h)$ is the spectator field potential and a prime indicates differentiation with respect to the field value. The eigenfunctions $\psi_n(h)$ form an orthonormal and complete basis, which can be used to determine the equilibrium probability distribution of the spectator field values as~\cite{Starobinsky:1986fx,Starobinsky:1994bd}
\begin{equation}
P_{\rm eq}(h) = \psi_0^2(h)\propto \exp\bigg\{-\f{8\pi^2}{3H^4}V(h)\bigg\}\,.\label{eq:prob}
\end{equation}
For a theory with $V(h)=(\lambda/4)h^4$ a convenient dimensionless variable is $x:=h(H/\lambda^{1/4})^{-1}$, which provides useful insights for practical calculations  we will frequently make use of: the region with $|x|\gtrsim1$ is exponentially suppressed and can often be ignored. This is visible in (\ref{eq:prob}) for $\psi_0$ and is also true for the higher order eigenfunctions.

A spectator field exhibits sizeable fluctuations in de Sitter space if the condition $V''(h)\ll H^2$ holds. Otherwise, the field begins to evolve classically according to its potential and quickly settles to its minimum value.

The computation of a generic temporal two-point correlation function proceeds by demarginalization of the two-field joint probability distribution function in terms of the equilibrium one-field distribution and the related conditional probability distribution (the transfer matrix of Ref.~\cite{Markkanen:2019kpv}). The expression for the temporal correlator is then extended to arbitrary two-point functions by means of the de Sitter invariance. For instance, the purely spatial correlators $G_f\left(\mathbf{x} ,\mathbf{x}'\right)=\braket{f( h(t, \mathbf{x})), f( h(t, \mathbf{x}'))} $ relevant for the present analysis are obtained via the eigenfunctions and eigenvalues of Eq.~\eqref{eq:EIGS} as
\begin{equation}
	G_f\left(\mathbf{x} ,\mathbf{x}'\right) = \sum_{n=0}^\infty f_n^2 \,(a_{\rm} \,r \, H)^{-\frac{2\Lambda_n}{H} }\,.
\end{equation}
As we can see, the correlation function depends on the comoving separation between the points $r:=\modu{\mathbf{x} - \mathbf{x}'}$ and the contribution of each eigenfunction is given by the coefficients 
\begin{equation}
	f_n:=\int\limits_{-\infty}^\infty \td h\, \psi_0(h)\, f(h)\,\psi_n(h)\,.
\end{equation}  
The power spectrum of $G_f\left(\mathbf{x} ,\mathbf{x}'\right)$ {is defined} via a Fourier transformation. It is often the case that one is interested only in the large scale limit, where the spectrum has the following form\footnote{For more complicated potentials such as the double well investigated in Ref.~\cite{Markkanen:2020bfc}, the first non-zero coefficient can remain subdominant until scales much larger than the ones relevant in cosmology.}
\begin{align}
{\cal P}_f(k)&=\frac{k^3}{2\pi^2}
\int d^3x e^{-i\mathbf{k}\cdot\mathbf{x}}\langle f(h(0)) f(h(\mathbf{x}))\rangle\simeq\f{2}{\pi}f^2_d\Gamma\bigg(2-2\f{\Lambda_d}{H}\bigg)\sin\bigg(\f{\Lambda_d\pi}{H}\bigg)\bigg(\f{k}{aH}\bigg)^{\f{2\Lambda_d}{H}}\nonumber \\&\simeq \f{2\Lambda_d}{H}f^2_d\bigg(\f{k}{aH}\bigg)^{\f{2\Lambda_d}{H}}+{\cal O}(\Lambda_d^2/H^2)\,,\label{eq:lP}
\end{align}
and the subscript `$d$' indicates the dominant contribution to be determined from Eq.~\eqref{eq:EIGS}.

Even though the calculation above is for de Sitter space, it is believed to be a good approximation for the inflationary period as long as the Hubble rate $H$ is slowly varying. The power spectrum at the end of the inflationary epoch is then obtained by setting $a=a_{\rm end}$ and $H_{\rm end}$ in Eq.~\eqref{eq:lP}, where a subscript `$\rm end$' indicates that the  quantity is to be evaluated at the end of inflation.

\section{Deriving the power spectrum with the \texorpdfstring{$\delta N $}{Lg} formalism}
\label{sec:dn}

The full power spectrum of curvature perturbations can be computed using the $\delta N$ formalism. In this method\footnote{ We refer the reader to Ref.~\cite{Ichikawa:2008ne} for the treatment of reheating modulation within the alternative mean field approach.}, at the leading order in spatial gradients, the evolution of coarse grained super-horizon regions is regulated by local Friedmann equations evaluated separately for each of these patches~\cite{Wands:2000dp,Starobinsky:1986fxa,Salopek:1990jq,Sasaki:1995aw,Sasaki:1998ug}.

For definiteness we assume that inflation is driven by a single scalar field $\phi$,
{the inflaton}, slowly rolling along its potential.  We furthermore assume that $h$ is the only light scalar spectator field present and that it remains energetically subdominant until it eventually thermalizes after reheating. Hence we  neglect the corresponding contribution in writing the Friedman equations that regulate the evolution of super-horizon regions. For the sake of the present discussion we assume an implicit dependence of reheating dynamics on $h$, writing for the corresponding energy density $\rho_{\rm reh}=\rho_{\rm reh}(h)$. The origin of such relation is analyzed in detail in the forthcoming section.

The expansion for each super-horizon region, from an initial time $t_{\rm in}$ during inflation to a final time with fixed reference energy $\rho_{\rm f}$ after reheating, can be quantified in the local number of $e$-folds as
\begin{align}
\label{nschematic}
N({\bf x}) 
=
\int\limits_{\rho_{\rm in}(\bar{\phi}({\bf x}))}^{\rho_{\rm end}}\frac{H}{\dot{\rho}}\td\rho
+
\int\limits_{\rho_{\rm end}}^{\rho_{\rm reh}(\bar{h}({\bf x}))}\frac{H}{\dot{\rho}}\td\rho
+
\int\limits_{\rho_{\rm reh}(\bar{h}({\bf x}))}^{\rho_{\rm f}}\frac{H}{\dot{\rho}}\td\rho ~.
\end{align}
Here $\bar{\phi}(\bf{x})$ and $\bar{h}(\bf{x})$ denote the local initial field values at $t_{\rm in}$, while $\rho_{\rm end}$ and $\rho_{\rm reh}$ are the values of the energy density at the end of inflation and reheating, respectively. We remark that $N$ depends on the spectator field value only through $\rho_{\rm reh}=\rho_{\rm reh}(\bar{h})$. 

In order to evaluate the above integrals, we use the leading order slow-roll approximation $3H\dot{\phi} = -V'(\phi)$ over the range $[t_{\rm in}, t_{\rm end}]$, and assume a perfect fluid equation of state with constant $w$ for the interval $[t_{\rm end}, t_{\rm reh}]$. As for the last term, which models the contribution after reheating, we assume a radiation dominated universe and thus obtain
\begin{align}
\label{nexplicit}
N({\bf x}) = 
- \int\limits_{\bar{\phi}({\bf x})}^{\phi_{\rm end}}\frac{1}{\sqrt{2\epsilon}M_{\rm Pl}}\td \phi
-\frac{1}{3(1+w)}\ln\frac{\rho_{\rm reh}(\bar{h}({\bf x}))}{\rho_{\rm end}}
-\frac{1}{4}\ln\frac{\rho_{\rm f}}{\rho_{\rm reh}(\bar{h}({\bf x}))}~.
\end{align}

The curvature perturbation on uniform density slices at super-horizon scales is therefore computed as  
\be
\zeta({\bf x}) := N({\bf x})-\langle N ({\bf x})\rangle\,,
\ee
where the gauge choice is imposed by setting the final energy density after reheating to a fixed reference value $\rho_{\rm f} = \langle \rho \rangle$, independent of ${\bf x}$. 
Using Eq.~\eqref{nexplicit} yields   
\be
\label{zetafull0}
\zeta({\bf x}) = \zeta_{\phi}({\bf x}) -
\frac{1-3w}{12(1+w)}\Big[\ln\rho_{\rm reh}(\bar{h}({\bf x}))-\Big\langle\ln\rho_{\rm reh}(\bar{h}({\bf x}))\Big\rangle\Big],
\ee
where the first term is the usual inflaton contribution. Notice that if the spectator contribution to local Friedmann equations is negligible, consistently with Eq.~\eqref{nexplicit}, there are no isocurvature perturbations present after reheating and $\zeta({\bf x})$ remains constant in time. 

By defining $\delta\rho_{\rm reh}({\bf x}) := \rho_{\rm reh}(\bar{h}({\bf x})) - \langle\rho_{\rm reh}(\bar{h}({\bf x}))\rangle$ and expanding to leading order in $\delta\rho_{\rm reh}/\langle\rho_{\rm reh}\rangle$, Eq.~\eqref{zetafull0} becomes
\begin{align}
\label{zetafull}
\zeta &\simeq \zeta_{\phi}({\bf x}) -
\frac{1-3w}{12(1+w)}\frac{\delta\rho_{\rm reh}({\bf x})}{\langle\rho_{\rm reh}\rangle}
\,.
\end{align}
The $\delta N$ expression for the curvature perturbation is by construction independent of the initial time $t_{\rm in}$, which labels a spatially flat hypersurface, as long as it is after the horizon exit of all modes of interest \cite{Lyth:2004gb}. Here we choose a time $t_{\rm in}$, at which $\bar{h}({\bf x})$ (and $\bar{\phi}({\bf x})$) are evaluated, just before the end of inflation\footnote{Setting $t_{\rm in}  = t_{\rm end}$ would define a uniform inflaton field gauge through $\epsilon(\phi_{\rm end}) =1$, therefore we choose the spatially flat slice $t_{\rm in} $ slightly before $t_{\rm end}$.}. 

Since the fields $\bar{h}({\bf x})$ and $\bar{\phi}({\bf x})$ are mutually uncorrelated, the spectrum of the curvature perturbation ${\cal P}_\zeta$ is given by the sum of the inflaton and spectator field power spectra: 
\begin{align}
{\cal P}_\zeta&=
{\cal P}_\zeta^{(\phi)}+\bigg(\frac{1-3w}{12(1+w)\langle\rho_{\rm reh}\rangle}\bigg)^2{\cal P}_{\delta\rho}\equiv{\cal P}_\zeta^{(\phi)}+{\cal P}_\zeta^{(h)}
~.\label{eq:ModP}
\end{align}
We observe that for $w=1/3$, corresponding to the inflaton oscillating in a quartic potential $V(\phi)\propto \phi^4$, the contribution from the spectator field vanishes identically: $\zeta = \zeta_{\phi}$. In fact, in this case every super-horizon patch transitions to a radiation dominated regime as the inflationary expansion concludes, regardless of the local value of $h$. However, for $w\neq 1/3$ the second term does not vanish and its contribution can be important.

\section{Indirect modulation}
\label{sec:toy2}
\subsection{A simple model with Yukawa interactions}
To demonstrate the mechanism of indirect modulation we consider a simple model consisting of an inflaton field $\phi$, a light spectator field $h$ and a fermion $\Psi$
\begin{equation}
	\label{eq:lag2}
	\LG = \frac{1}{2}(\pd\phi)^2 -\frac{1}{2}m^2_\phi\phi^2 + i \bar\Psi(\slashed{\pd} - m_\Psi)\Psi + \frac{1}{2}(\pd h)^2 -\frac{\lambda}{4} h^4 -y_\phi\bar\Psi\Psi\phi - y_h\bar\Psi\Psi h \,,
\end{equation}
where all coupling constants are assumed to be real and the considered inflaton potential is meant to describe solely the reheating dynamics that follows the initial expansion epoch. For the spectator field, we take a positive quartic coupling $\lambda>0$ that induces an effective field-dependent mass $\mu_h^2 = 3 \lambda h^2 >0$. The fermion $\Psi$ is also characterized by an effective mass $\mu_\Psi = m_\Psi + y_h h$, but no contribution from the inflaton is present since the field is rapidly oscillating around a vanishing field value. We will consider a regime where $y_h h \gg m_{\Psi}$, so we can safely take $\mu_\Psi \simeq y_h h$. 

Reheating proceeds via the perturbative decay of the inflaton into $\Psi$ pairs with a corresponding decay width given by
\begin{equation}
	\Gamma(h)=\frac{y_\phi^2  m_\phi}{8\pi}\left[1-\frac{(2y_hh)^2}{  m_\phi^2}\right]^{3/2}\,,
\end{equation}
however, the process is kinematically allowed only if $m_\phi > 2 y_h h$. In terms of the spectator field value, this defines the characteristic scale  
\begin{equation}
	\label{eq:hdec}
	\hkin:=\frac{m_\phi }{2 y_h},
\end{equation} 
such that the decay of the inflaton field, and thus reheating, can proceed only for $h< \hkin$.

\begin{figure}[t]
\centering
\includegraphics[width=.75\linewidth]{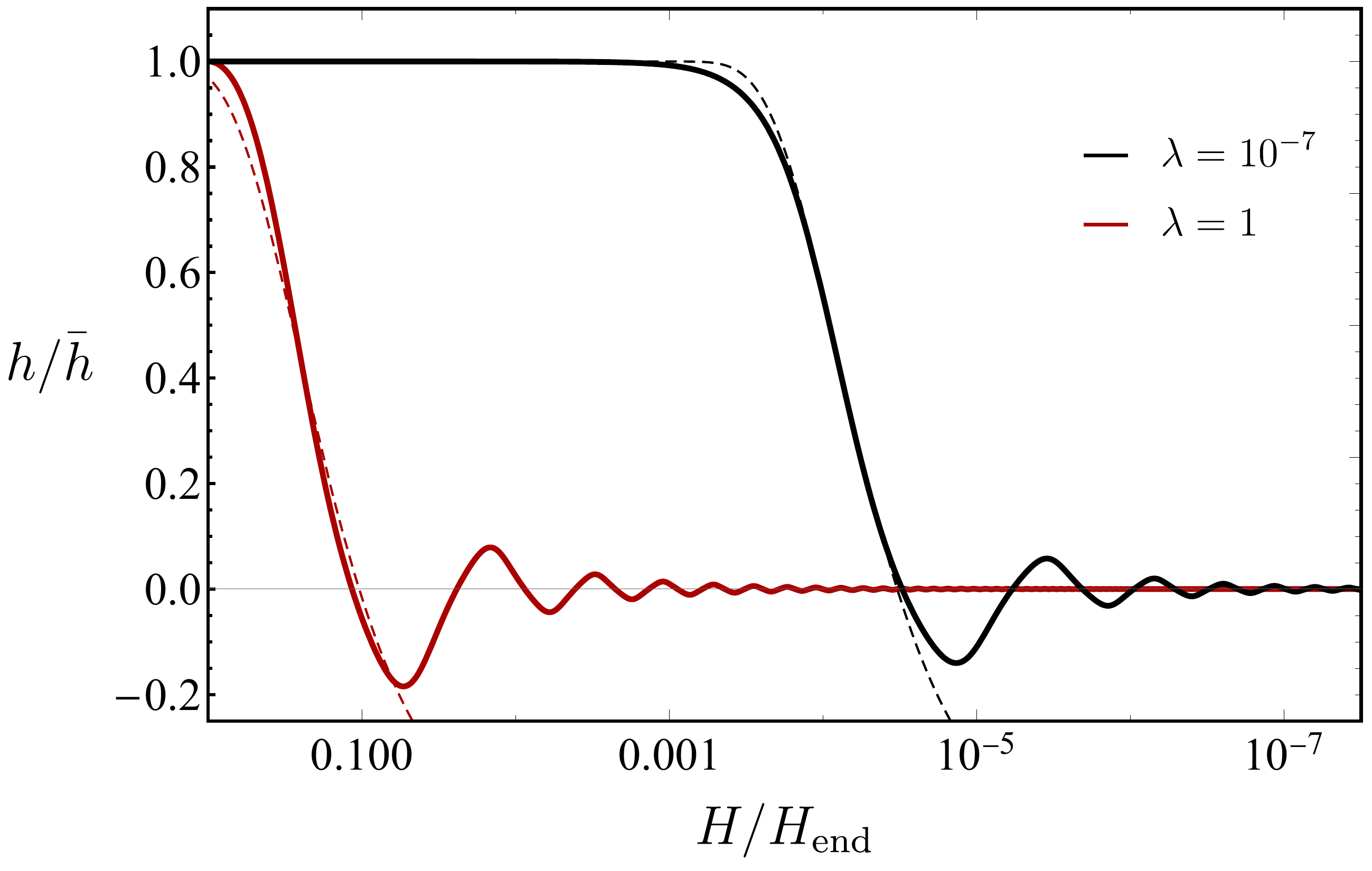}
\caption{The continuous lines show numerical solutions of the spectator field equation of motion in Eq.~\eqref{eq:eomeq}. The approximation in Eq.~\eqref{eq:pa} is indicated by the dashed lines. The red lines are for a quartic coupling $\lambda = 1$, while the black ones correspond to $\lambda = 10^{-7}$.}
\label{fig:1}
\end{figure}

Neglecting spatial gradients, the equation of motion of the spectator field reads
\begin{equation}
	\label{eq:eomeq}
	\ddot h + 3H\dot h + \lambda h^3=0\,, 
\end{equation}
where the background scales as dust $(w=0)$ when the inflaton potential during reheating is quadratic~\cite{Turner:1983he}. For the initial conditions $h=\bar{h}$ and $\dot{h}=0$  Eq.~(\ref{eq:eomeq}) has the approximate solution $h=\bar{h}$ until $H = \Hosc := \sqrt{3 \lambda} \bar h$, after it begins a series of damped oscillations as shown in Fig.~\ref{fig:1}.  

During inflation, the spectator field is fragmented into domains each characterized by a coherent \textit{local} value $\bar h(\mathbf{x})$ with the probability distribution in Eq.~\eqref{eq:prob} and a size determined by the two-point correlation function \cite{Markkanen:2020bfc}. The spatial gradients are typically small and can be ignored \cite{Starobinsky:1994bd}, so after inflation the evolution of each local value $\bar h(\mathbf{x})$ can be determined separately from the homogeneous equation of motion in Eq.~\eqref{eq:eomeq}. From now on for simplicity we will omit the ${\mathbf{x}}$-dependence from the initial value $\bar{h}$.

Given an initial condition $\bar{h}$ the evolution of the spectator during its first half oscillation can be faithfully tracked by using the following expression:
\begin{equation}
	\label{eq:pa}
	h=\bar h \left(1 - \frac 32 e^{-\frac{27}{4}\frac{H}{\sqrt{3\lambda}\bar{h}}}\right)\,,
\end{equation}   
indicated by the dot-dashed lines in Fig.~\ref{fig:1}. The approximation, which works exceptionally well across a large range of scales, was obtained by realising that the logistic function is often used to approximate the solution of similar differential equations~\cite{rohtua}. The coefficients are determined through a fit of the numerical result\footnote{Replacing the time derivatives in Eq.~\eqref{eq:eomeq} with derivatives with respect to $H$, the spectator field equation of motion in terms of $y:=h/\bar h$ becomes $ H^4 y'' + \frac{4}{27} H_{\rm osc}^2 y^3 = 0$, which provides an educated guess for the coefficient in the exponential.}.

The magnitude of the initial spectator field fluctuation scale relative to the scale $\hkin$ defines two scenarios:
\begin{itemize}
	\item $\bar h \leq\hkin$ -- no kinematic blocking. 
	In this case the spectator field cannot block the inflaton decay but can only modulate weakly through the mass dependence of the decay rate.
	
	\item $\bar h>\hkin$ -- kinematic blocking. \\In this case the large spectator field can initially completely block the inflaton decay, therefore  reheating can only occur after the spectator has relaxed below the critical value $\hkin$. 
\end{itemize}

For the modulation in the second case to take place, the spectator itself should of course not decay before the threshold $h_{\rm kin}$ is reached. The Lagrangian in Eq.~\eqref{eq:lag2} in principle also allows for the decay of the spectator field, however the process is generally negligible. The spectator field is  much lighter than the inflaton, so the same kinematic blocking factor forces the spectator field to decay after the inflaton. Quantitatively, for the model in Eq.~\eqref{eq:lag2}, the spectator field is stable if $4y_h^2/(3\lambda)>1$.

In the following we focus on the case where $\bar h>\hkin$, which we expect to result in a stronger modulation effect because of the presence of kinematic blocking. Due to fluctuations of the initial field value $\bar h$, the threshold is reached at different times at different locations.  We also simplify the computation of the power spectrum by neglecting potential additional modulation contributions from the region $\bar h<\hkin$. Because spectator field fluctuations are typically of the order ${\cal O}(H_{\rm end})$, Yukawa couplings of ${\cal O}(1)$ naturally result in an effective mass of the same order as the typical effective inflaton mass at reheating, $m_{\phi}\sim {\cal O}(H_{\rm end})$. Consequently, we expect the kinematic blocking of the inflaton decay to be present in a large part of the parameter space of the model. 

Modulated reheating through similar kinematic blocking has been investigated earlier in e.g. \cite{Fujita:2016vfj,Lu:2019tjj}. The crucial new ingredient here is the non-perturbative analysis of equilibrium spectator fields using the eigenvalue expansion of the stochastic approach discussed in Sec.~\ref{sec:spec101}. For example, as discussed in \cite{Markkanen:2019kpv} for equilibrium spectators in de Sitter space the standard mean field approach fails to give the correct spectral tilt and non-perturbative resummation techniques are required for their reliable analysis.  

\subsection{Power spectrum from indirect modulation}
A detailed calculation of the reheating dynamics is often a challenging problem, with possible non-perturbative aspects requiring the use of numerical methods~\cite{Felder:2000hq}.
However, for our purposes, it is sufficient to use the approximation where the inflaton instantaneously decays at the moment the decay channel opens during the first oscillation of the spectator field, valid for   
\begin{equation}
    \Gamma(h)\simeq\Gamma_0 := \frac{y^2_\phi m_\phi}{8\pi} \gg H_{\rm kin}\,,\label{eqhdec}
\end{equation}
where we have neglected the effective fermion mass by taking $m_\phi \gg y_h h$. In the above, $H_{\rm kin}$ is defined as the Hubble rate at the threshold $h=h_{\rm kin}$.

At this stage we need to find the explicit form of $H_{\rm kin}$. Inverting the Eq.~\eqref{eq:pa} evaluated at $h=\hkin$ then approximately yields the Hubble rate at the decay instant,
\begin{equation}
	H_{\rm kin}(\bar{h}) = 
	\frac{4}{27} \sqrt{3\lambda}\bar{h} \, \ln\left(\frac{3 \bar h}{2(\bar h - \hkin)}\right)\,.\label{eq:dec}
\end{equation}
As we can see, the Hubble rate at the inflaton decay time has spatial dependence induced by the spectator field value $\bar{h}$. We remark that the approximation used for $H_{\rm kin}$ requires a marginal modification of the modulation condition $h_{\rm kin}<\bar{h}$ into 
\begin{equation}
\frac{\hkin}{1-(3/2)\,e^{-27/4}}<\bar{h}
\,,\label{eq:rang}
\end{equation}
where the new lower bound $\approx1.0018h_{\rm kin}$ ensures that $H_{\rm kin}< H_{\rm osc}$, as required for consistency. 

As $\hkin$ sets the lower bound for a fluctuation to block the decay, in parameter regions characterized by a wide range of $\bar{h}$ values above this threshold, we may approximately assume $\bar h \gg \hkin$ and neglect $\hkin$ in Eq.~\eqref{eq:dec} in order to derive the condition for successful reheating in Eq.~\eqref{eqhdec}
\be
\f{\bar{h}}{H_{\rm end}/{\lambda^{1/4}}}< \frac{9 \sqrt{3} m_\phi y_\phi^2}{32
   \pi H_{\rm end} {\lambda^{1/4 }} \ln
   \left({3}/{2}\right)}\,,\label{eq:decb1}
\ee
written in terms of the dimensionless variable $\bar h(H_{\rm end}/\lambda^{1/4})^{-1}$. The further $\bar{h}$ is from the bound in Eq.~\eqref{eq:decb1} the better the condition in Eq.~\eqref{eqhdec} is satisfied. In order to avoid parameter regions where the approximation breaks down we will only include  cases where the right hand side of \eqref{eq:decb1} is larger than unity. As discussed in Sec.~\ref{sec:spec101} the bulk of the probability distribution in Eq.~\eqref{eq:prob} lies in the range $\bar h(H_{\rm end}/\lambda^{1/4})^{-1}\lesssim1$  so with this choice we include only cases where our approximation correctly captures the relevant field values.

Our approach is not suited for non-perturbative effects that possibly occur when the spectator field rapidly oscillates, so we restrict our analysis to the case in which the inflaton field decays during the first half oscillation cycle. By using Eq.~\eqref{eq:pa} we then compute
\begin{align}
\int\limits^{t(h=0)}_{t(h_{\rm kin})}\Gamma(h) \td t \approx\Gamma_0\Delta t &=\frac{3 \sqrt{3} m_\phi y_\phi^2}{16
   \pi  \bar{h} \sqrt{\lambda }}\bigg[\f{1}{\ln\left(\frac{3}{2}\right)}-\f{1}{\ln\left(\frac{3 \bar h}{2(\bar h - \hkin)}\right)}\bigg]\approx\frac{3 \sqrt{3} m_\phi^2 y_\phi^2}{32
   \pi\ln^2(3/2)  \bar{h}^2 y_h\sqrt{\lambda }}
   \,,
   \label{eq:decb2}
\end{align}
and ensure the robustness of our results by requiring that $\Gamma_0 \Delta t>1$, or
\be
\f{\bar{h}^2}{H_{\rm end}^2/\sqrt{\lambda}}< \frac{3 \sqrt{3} m_\phi^2 y_\phi^2}{32 \pi  H_{\rm end}^2\label{eq:decb3}
   y_h \ln ^2\left({3}/{2}\right)}\,.
\ee
Note that the right-hand side of the above equation is independent of $\lambda$, unlike the condition in Eq.~\eqref{eq:decb1}. When presenting our results in Sec.~\ref{sec:Results} we label as 'non-perturbative' all regions where the right-hand side of Eq.~(\ref{eq:decb3}) is smaller than unity.

When the bounds discussed in this section are satisfied, we can confidently use our approximations to calculate the spatial modulation of the reheating energy density $\rho_{\rm reh}$ with the help of sections and \ref{sec:spec101} and \ref{sec:dn}, induced through the dependence of the Hubble parameter at the inflaton decay on the spectator field value:
\begin{equation}
\label{eq:rhoreh}
	\rho_{\rm reh}(\bar{h}(\mathbf{x}))=3 \mpl^2 H_{\rm reh}^2 (\bar h(\mathbf{x}))\,, 
\end{equation}
where for clarity we explicitly write the $\mathbf{x}$-dependence of $\bar{h}$.


The results are summarized in the following expression for the Hubble rate at the reheating era 
\begin{equation}
\label{eq:hreh}
     H_{\rm reh}(\bar h(\mathbf{x})): = \begin{cases}
     H_{\rm kin} & \text{ if } \quad \dfrac{h_{\rm kin}}{1-(3/2)e^{-27/4}}\leq\bar{h}\leq\dfrac{9 \sqrt{3} m_\phi y_\phi^2}{32
   \pi \sqrt{\lambda{ }} \ln
   \left({3}/{2}\right)},
     \\
     \\
     \Gamma(h)\simeq\Gamma_0 & \text{elsewhere. } 
    \end{cases}
\end{equation}
The correlation function 
\begin{equation}
\bigg\langle\f{\delta\rho_{\rm reh}({0})}{\rho_{\rm reh}}\f{\delta\rho_{\rm reh}(\mathbf{x}) }{\rho_{\rm reh}}
 \bigg\rangle =
  \frac{\langle H^2_{\rm reh}(\bar h({0}))H^2_{\rm reh}(\bar h(\mathbf{x}))
  \rangle-\langle H^2_{\rm reh}\rangle^2}
  {\langle H^2_{\rm reh}\rangle^2}
\end{equation}
then yields the modulated component in the power spectrum of scalar perturbations determined by Eq.~\eqref{eq:ModP}.

In order to compare to the CMB observations, we compute the power spectrum of scalar perturbations at the pivot scale $k_*= a_* H_{*} $ through Eqs.~\eqref{eq:ModP} and \eqref{eq:lP} for a matter dominated regime ($w=0$), obtaining for the modulated component
\begin{align}
\label{eq:Ma}
{\cal P}^{(h)}_\zeta(k_*)=\f{f^2_2}{72\pi\langle H_{\rm reh}^2 \rangle^2}\Gamma\bigg(2-2\f{\Lambda_2}{\Hend}\bigg)\sin\bigg(\f{\Lambda_2\pi}{\Hend}\bigg)e^{\f{-2\Lambda_2}{\Hend}N_*}\,,
\end{align}
where the dominant contribution for a spectator field with quartic potential in de Sitter space is provided by the second eigenvalue $\Lambda_2=0.28938\sqrt{\lambda}\Hend$ which is also determines the spectral tilt in de Sitter as $d {\rm ln}{\cal P}_{\zeta}^{(h)}/d {\rm ln} k_*  = 2\Lambda_2/\Hend $~\cite{Markkanen:2019kpv}. In the above formula $N_* = {\rm ln}(a_{\rm end}/a_{*})$ is the number of $e$-folds at Hubble crossing of the pivot scale $k_*$, while the prefactor is determined by the following integrals: 
\begin{align}
\frac{f_2}{\langle H_{\rm reh}^2\rangle}\label{eq:resp}&=\f{\int \td \bar{h} \,\psi_0 H_{\rm reh}^2\psi_2}{\int \td \bar{h}\,\psi_0 H_{\rm reh}^2\psi_0}
=\frac{\int_0^{kc_1}\td x\, {\psi}_0 {\psi}_2
+\int^\infty_{c_2} \td x \, {\psi}_0 {\psi}_2
+\gamma\int_{kc_1}^{c_2} \td x\, {\psi}_0 {\psi}_2f(x)}{\int_0^{kc_1} \td x \,{\psi}_0 {\psi}_0+\int^\infty_{c_2} \td x \, {\psi}_0 {\psi}_0+\gamma\int_{kc_1}^{c_2} \td x \,{\psi}_0 {\psi}_0f(x)}\\ &\approx
\frac{\int_0^{kc_1}\td x\, {\psi}_0 {\psi}_2
+\gamma\int_{kc_1}^{\infty} \td x\, {\psi}_0 {\psi}_2f(x)}{\int_0^{kc_1} \td x \,{\psi}_0 {\psi}_0+\gamma\int_{kc_1}^{\infty} \td x \,{\psi}_0 {\psi}_0f(x)}\,;\quad\text{for}\quad c_2\gtrsim1\,,
\end{align}
where
\begin{align}
c_1&:= \f{m_\phi\lambda^{1/4}}{2\Hend y_h}\,;\quad  k:= \frac{1}{1-(3/2)e^{-27/4}}\,;\nonumber \\
\gamma&:= \f{1024\pi^2}{243}\f{\Hend^2\sqrt{\lambda}}{y_\phi^4m_\phi^2}\,;\quad c_2:= \frac{9 \sqrt{3} m_\phi y_\phi^2}{32
   \pi H_{\rm end} {\lambda^{1/4 }} \ln
   \left({3}/{2}\right)}
\,,\end{align}
and
\begin{align}
f(x):= x^2\log^2\bigg[\f{3x}{2(x-c_1)}\bigg]
\,.
\end{align}
The integrals in Eq.~(\ref{eq:resp}) are expressed in terms of the dimensionless variable $x:= \bar{h} (\Hend/\lambda^{1/4})^{-1}$
and the eigenfunctions must be evaluated numerically from Eq.~\eqref{eq:EIGS}. The term in the numerator proportional to $\gamma$ sources the modulation, which vanishes in the limit where the modulation window disappears, $kc_1\sim c_2$, because $\psi_0$ and $\psi_2$ are orthogonal. The effect also disappears for $c_1\gtrsim1$ since the probability of such high field values is exponentially suppressed. Similarly, since we consider only parameter regions where the right-hand side of Eq.~(\ref{eq:decb1}) is larger than unity i.e. $c_2 \gtrsim 1$, integrals extending from $c_2$ to infinity yield only negligible contributions.

When kinematic blocking occurs but the right-hand side of Eq.~(\ref{eq:decb3}) is smaller than unity strong modulation is expected. However, the inflaton decay process concludes only after the first oscillation cycle of the spectator field. The analysis of this regime needs to take into account possible resonant and non-perturbative effects~\cite{Enqvist:2015sua} that require methods beyond the scope of the present paper. For the sake of clarity we highlight these regions of the parameter space when presenting our results, expecting however that resonant and non-perturbative effect modify our results gradually as we enter these regions. At least for the cases we considered, the regions where the right-hand sides of both (\ref{eq:decb1}) and (\ref{eq:decb3}) are smaller than unity largely overlap. This can be traced back to an identical scaling in terms of $y_\phi$. 

The indirect modulation mechanism can be quite potent, leading to tight bounds when taking into account the observed amplitude of the CMB perturbations $\mathcal{P}_\zeta \simeq 2.1 \times 10^{-9}$~\cite{Akrami:2018odb}. As an example, for Yukawa couplings $y_\phi=y_h=1$, an inflaton mass $m_\phi=2H_{\rm end}$ and a quartic coupling $\lambda=10^{-5}$,  Eq.~\eqref{eq:Ma} yields ${\cal P}_\zeta^{(h)}\sim10^{-5}$ for the modulated component, demonstrating the importance of the effect.

\section{Indirect modulation in the Standard Model}
\label{sec:higgs}

We now demonstrate indirect modulation in the SM once non-vanishing neutrino masses are included. 

The observations of solar and atmospheric neutrinos unequivocally show the existence of two distinct mass scales in the neutrino sector, that source the measured flavour oscillation probabilities. It is then necessary to extend the SM particle content with (at least) two right-handed neutrino fields, which combine with the usual left-handed states to define the neutrino mass eigenstates. 

Since right-handed neutrinos are necessarily singlets under the SM gauge group it is possible to include a Majorana mass term in the Lagrangian allowing for a lepton-number violating interaction that may be tested experimentally.
The absence of such a signal would imply that neutrinos are Dirac fermions and acquire masses through the Higgs mechanism like the rest of the SM particles.

In order to accommodate non-vanishing neutrino masses and allowing for three right-handed neutrinos with corresponding Majorana masses, the SM Lagrangian is then extended to   
\begin{align}
	\LG& = \LG_{SM} + \frac12\left(\pd \phi\right)^2+(D_\mu H)^\dagger(D^\mu H) +i\bar{\ell}_L
	\slashed{D} {\ell}_L	+	i \brn \slashed{\pd} \rn\nonumber - \f12 m^2_\phi \phi^2-\lambda(H^\dagger H)^2\\&-\mu^2H^\dagger H+\Big\{-\frac12  \brnc M \rn
     - \f{Y_\phi}{2} \phi \, \brnc \rn - Y_h  \, \bar\ell_{\rm L}\rn \Tilde{H}  +\hc\!\Big\},\label{eq:LL}
\end{align}
where $\Tilde H = i \sigma_2 H^\dagger$,  with $H$ being the SM Higgs doublet, $\sigma_2$ is the second Pauli matrix and $\LG_{SM}$ includes the remaining SM Lagrangian terms not written explicitly. As before, the field $\phi$ represents the inflaton field which oscillates in a quadratic potential. We have left implicit the indices for the SM and right-handed neutrino generations, so each lepton doublet denoted by $\ell_{\rm L}= (\nu_{\rm L}, e_{\rm L})^T$ interacts with a right-handed neutrino $\rn$ through a complex matrix of Yukawa couplings $Y_h$. Similarly, the right-handed neutrinos couple to the inflaton $\phi$ through their own set of Yukawa couplings collected in $Y_\phi$. The Majorana mass  $M$ vanishes for Dirac neutrinos.

In order to show how the model detailed in Eq.~\eqref{eq:lag2} is straightforwardly recovered, we simplify the discussion by focusing on the case of one SM generation. We also neglect the Higgs mass term $\mu$ and value at the electroweak minimum since the typical scale of inflation is much higher than $v=246$ GeV.  In the unitary gauge, the Higgs doublet reads $H=(0, h/\sqrt 2)^T$ (so that $\Tilde H = (h/\sqrt{2}, 0)^T$) where $h$ is a real scalar field.

By gathering neutrino fields in arrays of definite chirality  
\begin{equation}
    n_L:=\begin{pmatrix}
        \nu_{\rm L} \\ \rnc
    \end{pmatrix}; \qquad 
    n_R:= \begin{pmatrix}
        \nu^c_{\rm L} \\ \rn
    \end{pmatrix},
\end{equation}
we can then write the mass terms in matrix form
\begin{equation}
\LG \supset
-\frac 12 \bar n_R
	 \operatorname{\mathbf{M}}
	 n_L + {\rm H.c.};
	 \qquad
	  \operatorname{\mathbf{M}}:=
	 \begin{pmatrix}
	0 & \frac{Y_h h}{\sqrt 2}\\ \frac{Y_h h}{\sqrt 2} & Y_\phi\phi+M
	\end{pmatrix}.
\end{equation}
The mass matrix $ \operatorname{\mathbf{M}}$ receives contributions from both the inflaton and the Higgs field. However, as the inflaton oscillates rapidly around the origin of its potential, the mass contribution averages to a vanishing value. The scale associated with the right-handed neutrinos Majorana mass is instead a free parameter of the model. The requirement of successful leptogenesis generally forces it to lie in a wide range that goes from a few KeV to the grand-unification scale, depending on the specifics of the chosen mechanism. In spite of that, the thermalization of right-handed neutrinos in the early Universe requires a reheating temperature that exceeds the Majorana mass scale, suggesting that generally $M<H_{\rm end}$. The fluctuations of spectator fields are usually comparable to the Hubble rate at the end of inflation such that the condition $y_h h \gg M$ is naturally satisfied. In this case, the mass matrix $ \operatorname{\mathbf{M}}$ can be approximated with

\begin{equation}
    \operatorname{\mathbf{M}}\xrightarrow[Y_h h \gg M]{}\begin{pmatrix}
	0 & \frac{Y_h h}{\sqrt 2}\\ \frac{Y_h h}{\sqrt 2} & 0
	\end{pmatrix}\,,
\end{equation}
and neutrinos recover the pure Dirac limit. By defining the Dirac neutrino field
\begin{equation}
    \Psi = \begin{pmatrix}
        \nu_L \\ N_R
    \end{pmatrix},
\end{equation}
in the limit $Y_h h \gg M$ the Lagrangian in Eq.~\eqref{eq:LL} reads
\begin{equation}
\label{eq:LLfinal}
	\LG \to \LG_{SM} + \f12\left(\pd \phi\right)^2 + \frac{1}{2}(\pd h)^2- \frac12 m^2_\phi \phi^2-\frac{\lambda}{4} h^4 + i \bar{\Psi}\left( \slashed{\pd} -\frac{Y_h h}{\sqrt 2}\right)\Psi
	 - \f{Y_\phi}{2}\phi\left(\overline{\Psi^c}P_R\Psi+\overline{\Psi}P_L\Psi^c\right)\,.
\end{equation}
The above Lagrangian then clearly resembles that of the model discussed in Sec.~\ref{sec:toy2}, after the Higgs field is identified with the spectator field.
An explicit computation of the decay width for the $\phi \to \Psi \Psi$ process 
at the massless limit, $\Gamma_0$ from Eq.~(\ref{eqhdec}) however differs
by a factor of $1/2$: the coupling of the inflaton to right-handed neutrinos forces it to decay via lepton number violating processes that yield pairs of neutrinos and antineutrinos, resulting in a reduction of $1/2$ in the available phase space. The conclusions of the previous section can then be applied to the SM case by simply identifying
\begin{equation}
    y_\phi \longleftrightarrow \frac{Y_\phi}{\sqrt 2}\quad\text{and}\quad   y_h \longleftrightarrow \frac{Y_h }{\sqrt 2}\,.\label{eq:lim}
\end{equation}

\section{Results}
\label{sec:Results}

\begin{figure}[ht]
\centering
\includegraphics[width=1\linewidth]{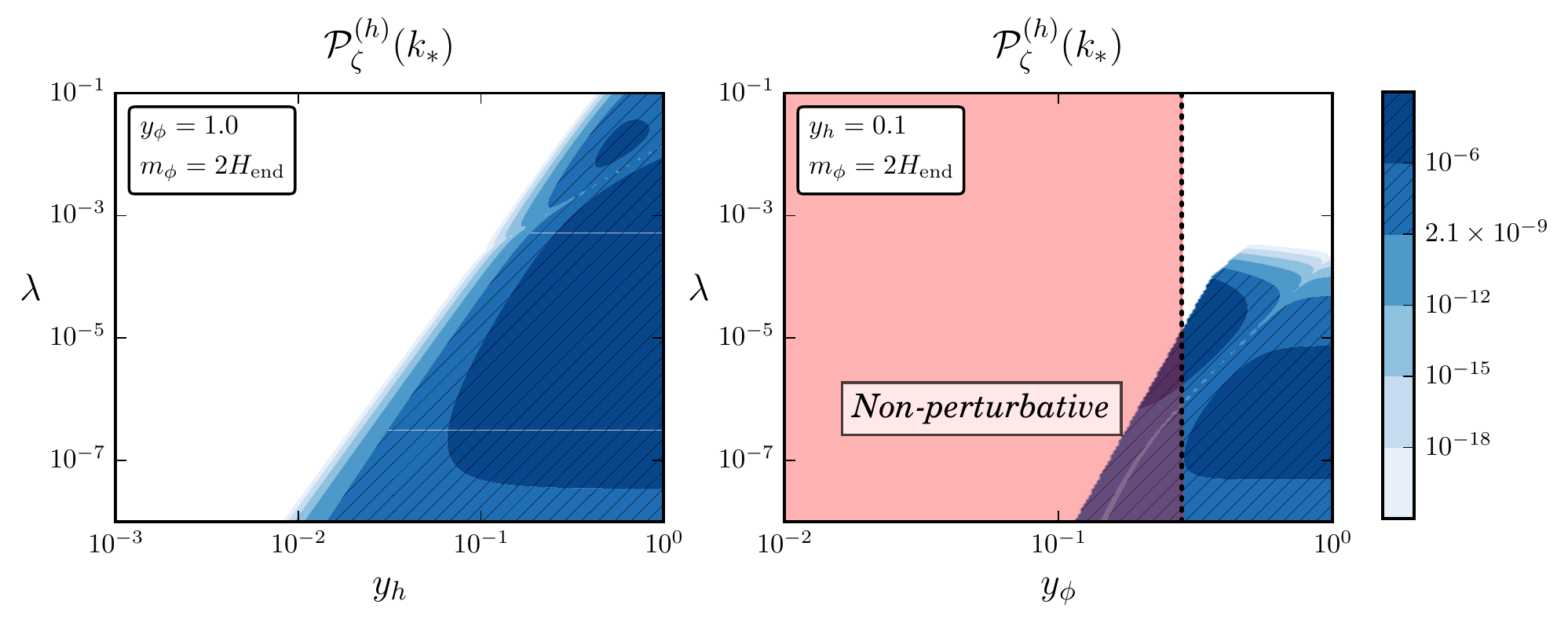}
\caption{The amplitude of perturbations at the pivot scale $N_*=60$ from indirect modulation. The hatched regions correspond to perturbations larger than ${\cal P}_\zeta(k_*)=2.1\times 10^{-9}$. The region where non-perturbative effects are important coming from Eq.~(\ref{eq:decb3}) is shaded red.} \label{fig:2}
\end{figure}

We describe now the results obtained for the parameters used in the simple model of Sec.~\ref{sec:toy2}. From the discussion above, it is clear that the same conclusions hold for the SM case if the spectator field is identified with the Higgs boson and the corresponding Yukawa couplings are properly rescaled. 
In Fig.~\ref{fig:2} we present two representative exclusion plots for the couplings, showing the regions where the spectrum of curvature perturbations from indirect modulation ${\cal P}_{\zeta}^{(h)}$ given by Eq.~\eqref{eq:Ma}, is larger than the observed value ${\cal P}_\zeta = 2.1\times 10^{-9}$ at the pivot scale $k_* = 0.05 {\rm Mpc}^{-1}$~\cite{Aghanim:2018eyx}. For the inflaton mass we take\footnote{For example, quadratic inflation yields $m_\phi = 1.4 \, H_{\rm end}$, while Starobinsky inflation results in $m_\phi = 3 \, H_{\rm end}$.} $m_\phi = 2H_{\rm end}$ and assume that the pivot scale exits the horizon $N_*=60$ $e$-folds before the end of inflation. In the right panel the non-perturbative region, where the right-hand side of Eq.~\eqref{eq:decb3} is smaller than unity, is shaded red. 

As one can see from Fig.~\ref{fig:2}, in some regions the amplitude of perturbations sourced by the indirect mechanism alone coincides with the one observed. However, interacting spectator fields in their vacuum state yield a blue-tilted spectrum \cite{Herranen:2013raa} and hence it is not possible to generate the observed CMB perturbations in this way.

\begin{figure}[ht]
\centering
\includegraphics[width=1\linewidth]{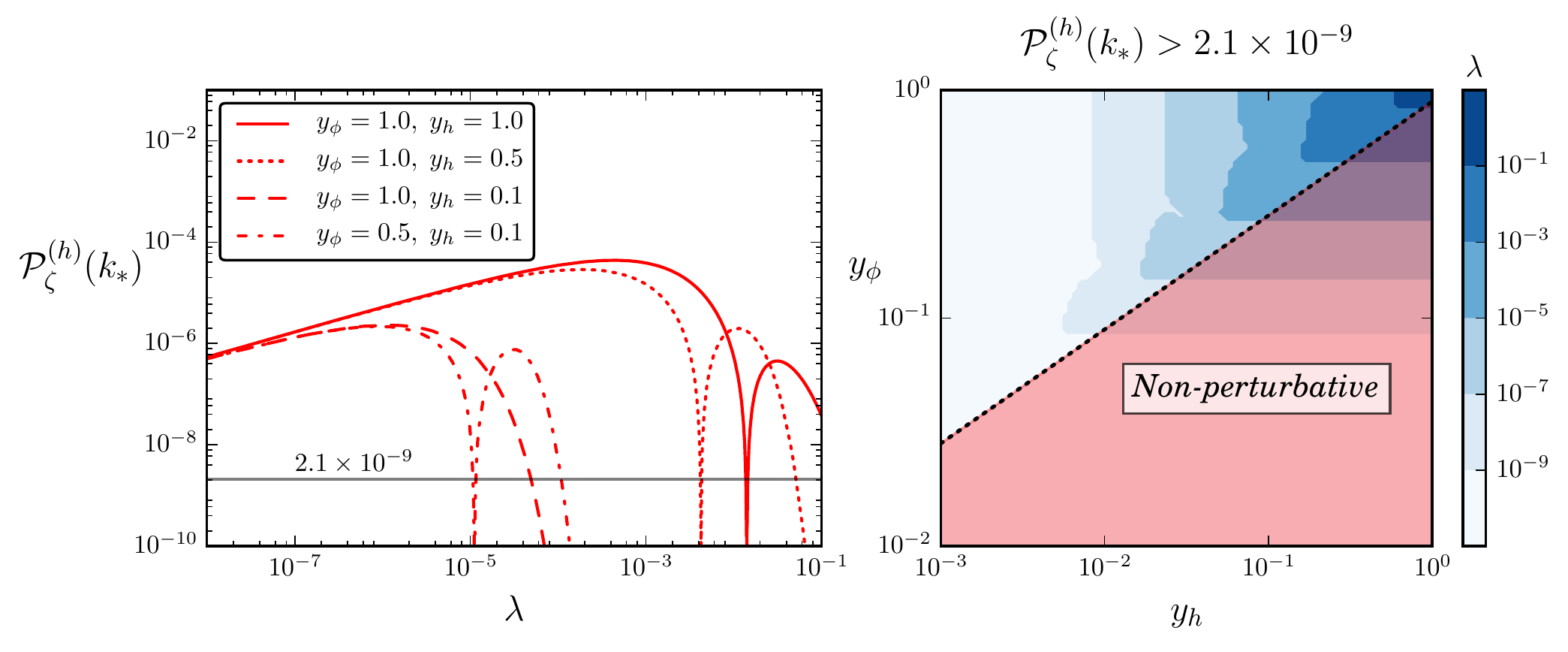}
\caption{(Left) The amplitude of perturbations at $N_*=60$ and with $m_\phi=2 H_{\rm end}$ for a set of choices for $(y_\phi,y_h)$. (Right) The parameter regions excluded by CMB observations with the same assumptions as in the left panel, where again areas where non-perturbative effects are important are shaded red.}\label{fig:3}  
\end{figure}

The left panel in Fig.~\ref{fig:3} highlights the behaviour of ${\cal P}_{\zeta}^{(h)}$ as a function of the spectator self coupling $\lambda$. The sharp feature seen here, and already visible in Fig.~\ref{fig:2}, is due to the sign changing eigenfunction contributions in the integrals in Eq.~(\ref{eq:resp}).
CMB constraints for the couplings of the Lagrangian \eqref{eq:lag2} from the indirect modulation are summarised in the right panel of Fig.~\ref{fig:3}. Coupling values above the $(y_h,y_{\phi},\lambda)$ surface depicted in the figure are excluded as the resulting ${\cal P}_{\zeta}^{(h)}$ is above the observed level. Decreasing the value of $\lambda$ increases the exclusion area as it yields larger spectator fluctuations and hence bigger ${\cal P}_{\zeta}^{(h)}$. Again, we have marked in the figure the non-perturbative regime where our simplified approximations of the decay process start to become inaccurate.

The theory described by Eq.~\eqref{eq:LL} that we used to study the CMB modulation due to the Higgs-like spectator field is nothing but the usual SM extended to accommodate the observed non-vanishing neutrino masses, leptogenesis and inflation. This Lagrangian has been extensively studied over the past decades and the parameter space that can successfully generate the light neutrino masses via the seesaw mechanism~\cite{Minkowski:1977sc} and the baryon asymmetry of the Universe via the leptogenesis~\cite{Fukugita:1986hr} has been identified. More recently, the discovery of the Higgs boson by the LHC experiments~\cite{Aad:2012tfa,Chatrchyan:2012ufa} revealed a  surprising feature of the SM -- the criticality of the Higgs potential. Namely, at high energy scales relevant for the seesaw mechanism, leptogenesis and inflation the Higgs boson quartic coupling $\lambda$ as well as its beta-function approximately vanish, implying a metastable vacuum for the SM~\cite{EliasMiro:2011aa, Degrassi:2012ry, Buttazzo:2013uya}. Using the most recent values of the SM parameters, $m_t=172.9 \pm 0.4$~GeV, $\alpha_s(M_Z) = 0.1179\pm 0.0010,$ $m_h=125.10\pm 0.14$~GeV, and two-loop renormalization group equations employed in~\cite{Gabrielli:2013hma}, we find that $\lambda$ runs negative at $\Lambda\sim 5\times 10^{11}$~GeV. This result is most sensitive to the actual top-quark mass. Reducing its value by $1\sigma$ results in 
$\Lambda\sim 5\times 10^{12}$~GeV, while within $3\sigma$ uncertainties the scale of Higgs criticality can be extended up to the Planck scale. Alternatively, the SM vacuum stability can be improved by coupling the Higgs boson to additional scalar singlet(s), such as the potentially allowed large Higgs coupling to inflaton, which prevents $\lambda$ from running negative~\cite{EliasMiro:2011aa}. Such studies are beyond the scope of the present paper and we adhere to the SM results in the following.

Interestingly, the allowed SM and seesaw parameter ranges are non-trivially restricted by the constraints presented in Figs.~\ref{fig:2} and~\ref{fig:3}. In particular, as seen in the right panel of Fig.~\ref{fig:3}, the CMB constraints from the indirect modulation imply a {\it lower} bound on the Higgs boson quartic coupling which depends on the right-handed neutrino Yukawas, all evaluated at the scale of inflation $H_{\rm end}$. The constraints are unaffected upon varying the value of $H_{\rm end}$ provided that the ratio $m_{\phi}/H_{\rm end}$ does not change and hold for any $H_{\rm end}\gtrsim M \lambda^{1/4}/y_h$. The first point follows from Eq. (\ref{eq:resp}) and the second from our assumption $y_h \bar{h} \gg M$, that lead to Eq.~\eqref{eq:LLfinal}, after setting $\bar{h}\lesssim H_{\rm end}/\lambda^{1/4}$. For $y_h<1$, the observed neutrino masses imply an upper bound
$M < 10^{13}$~GeV on the right-handed neutrino mass scale through the seesaw formula $m_\nu = (Y_h v)^2/(2 M)$. Successful leptogenesis, instead, requires $M > 10^9$~GeV~\cite{Davidson:2002qv} for non-degenerate right-handed neutrinos. For example, taking $y_h=Y_h/\sqrt{2}=0.03$ and $y_\phi=Y_\phi/\sqrt{2}=1.0$ (see Eq.~(\ref{eq:lim})), we estimate $10^9~\text{GeV} < M < 10^{10}$~GeV from  leptogenesis and neutrino masses, respectively,  and $\lambda(H_{\rm end})\gtrsim10^{-7}$ from Fig.~\ref{fig:3}. For the present central value of top-quark mass measurement, this is compatible with the SM running\footnote{Notice that for such values of the right-handed neutrino Yukawa couplings $y_h$ their contribution to the running of Higgs quartic coupling $\lambda$ is completely negligible.} of $\lambda$ provided that $H_{\rm end}\lesssim 5\cdot 10^{11}$~GeV, close to currently favoured scenarios such as the Starobinsky or Higgs inflation. On the other hand, for larger values of $y_h$ the bound $\lambda > \lambda_{\rm c}$ in Fig.~\ref{fig:3} rapidly becomes impossible to satisfy for the SM running of couplings, assuming a minimally coupled Higgs sector with no direct couplings to the spacetime curvature or to the inflaton.

\section{Summary and outlook}
\label{sec:summary}


In this work we have studied modulated reheating from spectator fields in their equilibrium with no direct couplings to the inflaton. The reliable analysis of equilibrium spectators requires non-pertrubative resummation which we implemented using the eigenvalue expansion in the stochastic approach. We found that the \textit{indirect modulation} from equilibrium spectators leads to significant production of curvature perturbations. Our results provide novel constraints for particle physics even in the absence of direct coupling between the inflaton and the spectator field. 


To set up the formalism, we focused on a particular phenomenological setting where the inflaton and the spectator couple to the same fermion field via two separate Yukawa terms. The effective fermion mass varies due to the spectator fluctuations, and for large enough fluctuations the inflaton decay is blocked until the spectator field falls below a kinematic threshold. Consequently, the reheating completes at different times in different locations of the universe sourcing curvature perturbation. We computed the spectrum of curvature perturbations from the indirect modulation with equilibrium spectators using the stochastic formalism combined with the $\delta N$ approach. We find that for Yukawas close to unity, the spectrum of the curvature perturbations exceeds the observed level by four orders of magnitude. Notably, the perturbation amplitude does not directly depend on the scale of inflation. 

As a concrete example, we studied the indirect modulation in the Standard Model extended by right-handed neutrinos and a singlet inflaton, assuming a minimally coupled Higgs sector with a vanishing coupling to spatial curvature and no direct couplings to the inflaton. In this setup, the Higgs is a light spectator and modulates the inflaton decay through the neutrino masses. The main results of our analysis are shown in Figs.~\ref{fig:2} and~\ref{fig:3} which constrain the most natural parameter space of the seesaw mechanism and leptogenesis that is commonly considered in phenomenological studies. Requiring that perturbations from the modulation do not exceed the observed CMB amplitude sets a {\it lower} bound on the Higgs quartic coupling $\lambda(H)$ at the scale of inflation. The bound shown in Fig.~\ref{fig:3} strongly depends on the right-handed neutrino Yukawa couplings $y_{h}$ and $y_{\phi}$. For $y_{h} = 0.03$ and $y_\phi=1.0$, the bound is compatible with the SM and $H< 10^{11-12}$~GeV but for $y_h$ of order unity the constraint on $\lambda(H)$ can no longer be satisfied assuming the SM running and the minimally coupled Higgs sector. In conclusion, our results constrain and specify the high-energy parameters of the SM, neutrino physics and inflation for the most natural and interesting values of the relevant parameters. We reiterate that the constraints apply when the Higgs is a light spectator during inflation which implies that its non-minimal coupling and possible couplings to the inflaton are assumed to be small.

An interesting question is whether the indirect mechanism in the SM with right handed neutrinos could be responsible for the observed curvature perturbation. From our results it is clear that the observed amplitude of curvature perturbations can be obtained through indirect modulation. However, reproducing the correct spectral tilt requires a modification of the setup, which we will address in a forthcoming paper~\cite{our:2020}. 

\section*{Acknowledgements}
This work was supported by the Estonian Research Council grants PRG356, PRG803, MOBTT86, MOBJD323, MOBJD381, MOBTT5 and by the EU through the European Regional Development Fund CoE program TK133 ``The Dark Side of the Universe".
This project has received funding
from the European Union’s Horizon 2020
research and innovation programme under the Marie Skłodowska-Curie grant agreement No.~786564.
{AR was funded by the UK Science and Technology Facilities Council grant ST/P000762/1 and Institute for Particle
Physics Phenomenology Associateship.}


\bibliography{biblio}

\end{document}